\documentclass[lettersize,journal]{IEEEtran}
\usepackage{amsmath,amsfonts}
\usepackage{array}
\usepackage[caption=false,font=normalsize,labelfont=sf,textfont=sf]{subfig}
\usepackage{textcomp}
\usepackage{stfloats}
\usepackage{url}
\usepackage{verbatim}
\usepackage{graphicx}

\usepackage{cite}
\usepackage{amsmath,amssymb,amsfonts}
\usepackage{xcolor}

\usepackage{makecell}
\usepackage{multirow}
\usepackage{threeparttable}
\usepackage{booktabs}
\usepackage{soul}
\usepackage[ruled,linesnumbered,lined,noend]{algorithm2e}
\usepackage{caption}

\newcommand{\Sec}[1]{Sec.~\ref{#1}}
\newcommand{\Tbl}[1]{Tbl.~\ref{#1}}
\newcommand{\Alg}[1]{Alg.~\ref{#1}}
\newcommand{\Fig}[1]{Fig.~\ref{#1}}
\newcommand{\Equ}[1]{Equ.~\ref{#1}}
\newcommand{\tit}[1]{\textit{#1}}
\newcommand{\ttt}[1]{\texttt{#1}}
\newcommand{\tbf}[1]{\textbf{#1}}
\newcommand{\algit}[1]{$\mathit{#1}$}

\newcommand{\proj}{\textsc{SwitchBlade}}

\newcommand{\blue}[1]{\textcolor[rgb]{0,0,1}{#1}}

\renewcommand{\paragraph}[1]{\vspace{2pt}\noindent\textbf{#1}\hspace*{.05cm}}

\graphicspath{ {./fig/} }

\begin{document}

\title{Accelerating Generic Graph Neural Networks via Architecture, Compiler, Partition Method Co-Design}

\author{Shuwen Lu, Zhihui Zhang, Cong Guo, Jingwen Leng~\IEEEmembership{Member,~IEEE,} Yangjie Zhou, Minyi Guo~\IEEEmembership{Fellow,~IEEE}
\thanks{The authors are with the Department of Computer Science and Engineering of Shanghai Jiao Tong University, Shanghai 201100, China.}
\thanks{Email:\{lushuwen, zhihui.zhang, guocong, leng-jw, yj\_zhou\}@sjtu.edu.cn, guo-my@cs.sjtu.edu.cn.}
\thanks{The first two authors contributed equally to this work. Zhihui Zhang, Jingwen Leng and Minyi Guo are corresponding authors of this paper.}
}

\markboth{Journal of \LaTeX\ Class Files,~Vol.~14, No.~8, August~2021}%
{Shell \MakeLowercase{\textit{et al.}}: A Sample Article Using IEEEtran.cls for IEEE Journals}


\maketitle

\begin{abstract}
	Graph neural networks (GNNs) have shown significant accuracy improvements in a variety of graph learning domains, sparking considerable research interest. 
To translate these accuracy improvements into practical applications, it is essential to develop high-performance and efficient hardware acceleration for GNN models. 
However, designing GNN accelerators faces two fundamental challenges: the high bandwidth requirement of GNN models and the diversity of GNN models.
Previous works have addressed the first challenge by using more expensive memory interfaces to achieve higher bandwidth. 
For the second challenge, existing works either support specific GNN models or have generic designs with poor hardware utilization.

In this work, we tackle both challenges simultaneously. 
First, we identify a new type of partition-level operator fusion, which we utilize to internally reduce the high bandwidth requirement of GNNs. 
Next, we introduce partition-level multi-threading to schedule the concurrent processing of graph partitions, utilizing different hardware resources. 
To further reduce the extra on-chip memory required by multi-threading, we propose fine-grained graph partitioning to generate denser graph partitions. 
Importantly, these three methods make no assumptions about the targeted GNN models, addressing the challenge of model variety.
We implement these methods in a framework called \proj{}, consisting of a compiler, a graph partitioner, and a hardware accelerator. 
Our evaluation demonstrates that \proj{} achieves an average speedup of $1.85\times$ and energy savings of $19.03\times$ compared to the NVIDIA V100 GPU. 
Additionally, \proj{} delivers performance comparable to state-of-the-art specialized accelerators.

\end{abstract}


\section{Introduction}
\label{sec:intro}

\IEEEPARstart{G}{raph} neural networks (GNNs) have gained significant momentum as researchers have begun to integrate the concept of \textit{graphs} into deep learning (DL)~\cite{Survey:GNN:Alg}.
By merging the end-to-end hierarchical learning capabilities of DL with the structural representation power of graphs, GNNs have achieved improved accuracy across various domains, including molecular science~\cite{Molecule}, recommendation systems~\cite{PinSage}, and transportation~\cite{transportation}.
To translate the algorithmic advancements of GNNs into practical applications, effective and efficient execution is crucial~\cite{NeuGraph}.
However, general-purpose processors, such as CPUs and GPUs, struggle with performance and energy inefficiency when executing graph-related operations~\cite{HyGCN-CAL,SGA-CAL}.
As a result, GNN-dedicated accelerators have been proposed.

Numerous efforts have focused on accelerating one of the most popular GNN models, Graph Convolution Networks (GCNs)~\cite{GCN,GraphSAGE}.
GCNs comprise two primary operators: the sparse graph adjacency matrix and the dense vertex feature matrix, which are typically organized into a two-stage computation.
Based on this formulation, various designs, including inter- and intra-stage optimizations, have been proposed to achieve high performance~\cite{HyGCN,GraphACT,GReTA,GRIP,AWB-GCN,I-GCN,GCNAX}.

{\renewcommand{\arraystretch}{1}
	\begin{table}[t]
		\centering
		
		\caption{Operations in popular GNN models.}
		\label{tbl:gnn-example}    
		\resizebox{\linewidth}{!}{
			\begin{threeparttable}
				\setlength\tabcolsep{2pt}
				\begin{tabular}{ccc} 	\toprule
					\textbf{Model} &
					\textbf{Aggregation ($a_i$)} & \textbf{Combination ($h^{l+1}_i)$} \\ \midrule
					GCN \cite{GCN} & $\sum_{j\in N(i)}h_j^l d^{-1/2}_{j}$ 
					&	\texttt{ReLU}$(d^{-1/2}_{i} W^l a_i)$ \\ \midrule
					GAT \cite{GAT} & $\sum_{j\in N(i)}\alpha_{ij} W^l h_j^l $ 
					&	\texttt{ReLU}$(a_i)$ \\ \midrule
					SAGE-Pool\cite{GraphSAGE} & $\max_{j\in N(i)}(W_{pool}^lh_j^l+b)$ 
					& \texttt{ReLU}$(W^l (h^l_i~||~a_i))$\\ \midrule
					GG-NN \cite{GGNN} & $\sum_{j\in N(i)}(W^l h_{j}^{l} + b)$ 
					& \texttt{GRU}$(h^l_i, a_i)$\\  \bottomrule
					
				\end{tabular}
			
				\begin{tablenotes}[flushleft]
				\item
					where $\alpha_{ij}$ is the attention coefficient whose calculation is omitted in this table, $||$ is matrix concatenation, \texttt{MLP} is Multi-Layer Perceptron~\cite{MLP}, and \texttt{GRU} is Gated Recurrent Unit~\cite{GRU}.
					
				\end{tablenotes}
			\end{threeparttable}
		}
		

	\end{table}
}

GNNs, however, encompass a broad category that varies in both the number and combination of operators~\cite{gnn-benchmark}.
As shown in \Tbl{tbl:gnn-example}, several popular GNN models exhibit quite different characteristics in each stage, despite being divisible into two-stage forms.
Thus, prior GCN-specific accelerators may not offer the same performance and efficiency when executing other models.
Additionally, designing dedicated accelerators for every GNN model is impractical due to the high hardware development costs and the vast GNN model space.
As a result, a generic solution is desirable for the diverse range of GNN models.
Though uniform architectures have been proposed to address these challenges~\cite{DAC:AutenT020,EnGN}, they suffer from long-distance data movement~\cite{ReGraphX}, leading to increased latency and energy consumption.

Another challenge in GNN acceleration is the high bandwidth requirement.
Since GNNs combine deep learning with graph processing~\cite{Survey:GNN:Alg}, they inherit the characteristics of large feature maps and poor data locality~\cite{convert}.
Both characteristics necessitate substantial reads and writes to off-chip DRAM.
Moreover, current GNN models are primarily executed in an operator-by-operator paradigm, where all operators read and write to DRAM, and modern GNN models typically comprise ten or more operators in one layer~\cite{graphiler}.
As a result, the current GNN model execution leads to massive off-chip data access~\cite{tvm}.
Such high requirements make hardware bandwidth a potential bottleneck in GNN execution.
Existing works addressing GNN acceleration satisfy this high bandwidth requirement through emerging yet costly techniques, such as Processing-In-Memory (PIM)~\cite{gnn:accel:pim,ReGraphX}.
This work aims to provide a more cost-effective solution to overcome this challenge.

Though considerable effort has been expended, none of the prior works address both variety and bandwidth challenges simultaneously.
To fill this gap, this work seeks to tackle both challenges within a single system.
Our core idea is to optimize GNN bandwidth requirements without making any assumptions about the targeted GNN model structure.
Guided by this idea, we propose three generic methods.

The first method is partition-level operator fusion (PLOF).
Operator fusion has been proven to effectively mitigate the high bandwidth requirement challenge for conventional neural networks~\cite{tvm}.
However, previous efforts have neglected the irregular graph-traversal-based operators (GTRs), which are central to GNNs.
In this work, we propose a new graph partition-level operator fusion that fuses operators in arbitrary GNN models into three phases to alleviate high bandwidth requirements.
This is based on two observations.
First, all operators can be reorganized into a three-phase paradigm by borrowing the programming model of traditional graph processing~\cite{GAS}.
Second, graphs are typically partitioned into smaller components for on-chip loading and processing due to their large data size.
As a result, we can transfer data only at phase boundaries rather than operator boundaries.

The second method is shard-level multi-threading (SLMT).
Although operators can be fused together to reduce memory footprint, they are still executed sequentially.
Since each operator utilizes only one part of hardware resources, including bandwidth, resource utilization for other parts can be low during that operator's execution.
To enhance the utilization of different hardware resources simultaneously, we introduce shard-level multi-threading, where different shards of the graph are assigned to different hardware units.
This parallelizes the execution of multiple fused phases across different shards, allowing for more efficient use of hardware resources and further reducing the memory bandwidth requirements.

The third method is fine-grained graph partitioning (FGGP) on host.
The above two methods deploying multiple threads for concurrent processing improves hardware utilization yet increases the on-chip memory pressure.
To mitigate the memory-concurrency contention, we further propose fine-grained graph partitioning running on host device.
FGGP generates denser partitions to store more effective data under the same memory budget so as to improve the graph data reuse and reduce the bandwidth requirement.

All three proposed methods enhance hardware performance for GNN acceleration without making any assumptions about the GNN model structure. 
As a result, they offer a fresh perspective to design more flexible architectures that achieve both excellent applicability and high performance.

\begin{figure}[t]
	\centering
	\includegraphics*[trim=0.cm 0.cm 0.cm 0.cm,width=1\linewidth]{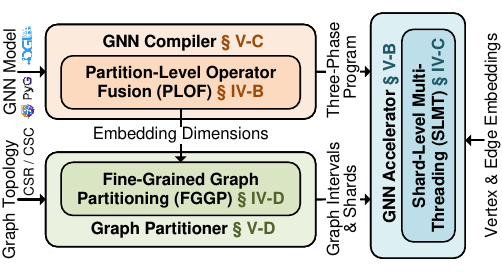}
	\caption{The workflow of \proj{}.}
	\label{fig:workflow}
\end{figure}

We propose a systematic framework, \proj{}, which comprises a compiler, a graph partitioner, and a hardware accelerator to implement the above methods, as illustrated in \Fig{fig:workflow}. The compiler is responsible for implementing PLOF.
It maps GNN models written in high-level frameworks such as DGL and PyG into our PLOF phases, which are expressed in the instruction set architecture (ISA) of our hardware accelerator.
Additionally, the compiler provides the model information to our graph partitioner, which implements FGGP and generates graph partitions.
The accelerator ultimately executes the PLOF phases and processes the graph partitions concurrently using SLMT.

We evaluate \proj{} in comparison to the NVIDIA V100 GPU and demonstrate that it achieves an average speedup of 1.85$\times$ and energy savings of 19.03$\times$ across a diverse range of GNN models. 
Furthermore, \proj{} attains comparable or even superior performance on GCN models when compared to state-of-the-art GCN accelerators like HyGCN~\cite{HyGCN}, but with significantly greater flexibility.

The main contributions of this work are as follows:
\begin{itemize}
	
	\item We propose a set of generic methods that make no assumptions about the underlying GNNs, addressing the bandwidth and variety challenges of generic GNN acceleration. These methods span algorithmic, software, and hardware aspects.
	
	\item We develop \proj{}, a comprehensive full-stack framework that implements the proposed three methods. The framework consists of a compiler, a graph partitioner, and a hardware accelerator.
	
	\item We evaluate the performance of \proj{} and demonstrate its effectiveness by achieving a 1.85$\times$ speedup and 19.03$\times$ energy savings compared to the NVIDIA V100 GPU. Furthermore, \proj{} exhibits comparable performance against a prior state-of-the-art GCN accelerator, showcasing its flexibility and adaptability.
	
\end{itemize}

\section{Background}
\label{sec:background}

In this section, we provide an overview of Graph Neural Networks (GNNs) and the dual-sliding-window graph partitioning method as the foundation for \proj{}.

\subsection{Graph Neural Networks}
\label{subsec:background:gnn}

Graph Neural Networks (GNNs) combine the power of deep learning with traditional graph processing, leading to improved accuracy in a wide range of domains that depend on graph structures, such as molecular science\cite{Molecule}, recommendation systems\cite{PinSage}, and transportation~\cite{transportation}. 
This improvement is achieved by replacing prior hand-crafted or intuition-based methods (e.g., node2vec~\cite{node2vec}) with end-to-end learning capabilities.

Similar to conventional deep neural networks (DNNs), GNNs consist of layers. 
A layer $l$ takes the vertex and edge embedding matrix as input, along with the adjacency matrix representing the graph structure, and produces a new embedding matrix for layer $l+1$\cite{Survey:GNN:Alg}. 
The primary distinction between GNNs and DNNs lies in the graph-traversal operators, which exhibit irregular computation and memory access patterns\cite{HyGCN-CAL}. 
We now introduce the primitive operators and their role in various GNN models.

\paragraph{Primitive Operators.}
Each GNN layer typically comprises two types of primitive operators: the \underline{g}raph-\underline{tr}aversal operators (\texttt{GTRs}, detailed below) and neural network operators. 
The latter can be further categorized as either \underline{d}ense \underline{m}atrix \underline{m}ultiplication operators (\texttt{DMMs})~\cite{MLP} and \underline{el}ement-\underline{w}ise operators (\texttt{ELWs}) such as \texttt{ADD}, \texttt{EXP}, and \texttt{ReLU}. 
The combination of these three operator types covers all forms of GNN computation. 
Most GNN libraries, including popular ones like DGL~\cite{dgl} and PyG~\cite{pyg}, support GNN programming by providing efficient implementations for these three primitive operators.

\texttt{GTRs} can be generalized into two types: \texttt{ScatterOp} and \texttt{GatherOp}. 
They are considered as \emph{vectorized} graph propagation operators in traditional graph processing models (e.g., GAS~\cite{GAS}). 
The \texttt{ScatterOp} distributes the embedding of each vertex to its outgoing (or incoming) edges, while the \texttt{GatherOp} collects the embeddings of all incoming (or outgoing) edges of each vertex and reduces them into a fixed-length vector using a \emph{reduction} function. 
Examples of reduction functions include $max$, $sum$, and $mean$, which align vertices' embeddings for subsequent operators, as different vertices may have varying edge numbers.

\begin{figure}[t]
	\centering
	\includegraphics*[trim=0.cm 0.cm 0.cm 0.cm,width=1\linewidth]{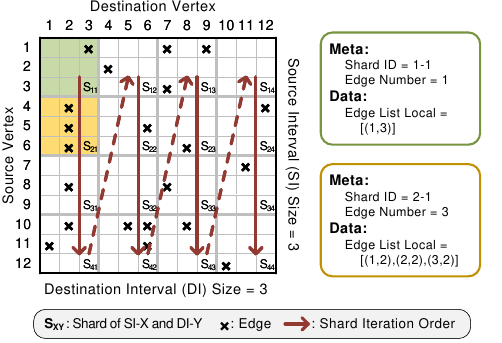}
	\caption{Dual-Sliding-Window Graph Partitioning (DSW-GP) adopted in \proj{}. 
		The red arrow indicates the shard processing order to traverse the while graph.}
	\label{fig:dsw-gp}
\end{figure}

\subsection{Dual-Sliding-Window Graph Partitioning}
\label{subsec:dsw-gp}

Traditional graph processing~\cite{gunrock} often has a memory footprint that can reach tens of gigabytes, including vertex and edge features, which exceeds both on-chip and off-chip memory capacities. 
Numerous graph partitioning techniques have been studied to reduce memory footprints~\cite{METIS, large_graph_partition}.

One notable approach is the dual-sliding-window-based graph partitioning (DSW-GP) method~\cite{GridGraph}. 
This method is popular due to its regular memory access patterns and has been adopted by several graph processing systems~\cite{GraphR}. 
DSW-GP first divides vertices into disjoint destination \textit{intervals} and then creates smaller \textit{shards} containing edges under each destination interval. 
\Fig{fig:dsw-gp} illustrates an example graph partitioned into 4 destination intervals and 16 shards. 
During partitioning, it is ensured that each shard can fit into the targeted memory space. 
During computation, operations are performed on each interval and shard instead of the entire graph, with shards being iterated interval-wise—referred to as window sliding.

GNNs, as an extension of traditional graph processing, also demand a large amount of memory concerning both graph scale and embedding size. 
To reduce memory footprint in GNN acceleration, we consider employing DSW-GP in this work.

\begin{algorithm}[t]
	\caption{\tbf{\tit{DSW-GP}}}
	\label{alg:gp:bg}
	
	\DontPrintSemicolon
	\SetKwInput{Input}{Input}
	\SetKwInput{Output}{Output}
	\Input{\algit{G}, input graph; \algit{M}, targeted GNN model.}
	\Output{\algit{S}, the resulted shards.}
	
	\algit{shardHeight \leftarrow} \tit{\tbf{calShardHeight}}\algit{(G,M)};\\
	\algit{shardNumInterval \leftarrow G.vTotal~/~shardHeight};\\
	
	\algit{I \leftarrow} \tit{\tbf{initIntervals}}\algit{(G,M)};\\
	\For{{\normalfont each}~\algit{i}~{\normalfont in}~\algit{I}~}{
		\For{\algit{sidx}~{\normalfont in~range}\algit{(shardNumInterval)}~} {
			\algit{srcBegin\leftarrow sidx\times shardHeight};\\
			\algit{srcEnd\leftarrow srcBegin + shardHeight};\\
			\If{\algit{srcEnd > G.vTotal}} {
				\algit{srcEnd\leftarrow G.vTotal};\\
			}
			\algit{s \leftarrow} \tit{\tbf{initShard}}\algit{(i)};\\
			\tit{\tbf{setShardSource}}\algit{(s, srcBegin, srcEnd)};\\
			\tit{\tbf{finalizeShard}}\algit{(i,s)};\\
		}
	}
	\tit{\tbf{finalizeInterval}}\algit{(i)};\\
\end{algorithm}

\section{Motivation}
\label{sec:motivation}

In this section, we discuss the motivation behind designing our \proj{} GNN accelerator framework. 
We first analyze the characteristics of GNNs to identify two primary challenges in GNN acceleration. 
Next, we review existing solutions and find that none of them address both challenges simultaneously. 
Finally, we present the goal of our \proj{} framework, inspired by the above insights.

\subsection{Challenges of GNN Acceleration}

By analyzing GNN characteristics, we identify two significant challenges in GNN acceleration:

\paragraph{High model variety.}
GNNs exhibit a high degree of variety in model structure. 
Unlike graph processing and deep learning, which are dominated by GEMM/CONV and SpMV/SpMM operations~\cite{Gamma}, GNNs do not have a single computational hotspot. 
Instead, they are more ad-hoc, depending on the targeted applications~\cite{Survey:GNN:Alg}. 
This characteristic leads to the challenge of accommodating the diverse GNN model structures when designing accelerators.

\paragraph{High bandwidth demand.}
Previous studies have shown that graph processing and deep learning applications are often limited by off-chip memory bandwidth~\cite{tvm}. 
While deep learning achieves high bandwidth utilization when transferring large amounts of weights and feature maps, graph processing experiences lower bandwidth utilization due to irregularity or sparsity in the graphs. 
As a combination of both graph processing and deep learning~\cite{Survey:GNN:Alg}, GNNs demand even higher bandwidth, exceeding today's bandwidth capacity.

\subsection{Limitations of Existing Solutions}

Although numerous GNN accelerators have been proposed in recent years, achieving significant performance improvements compared to general-purpose architectures, none of them fully address the two challenges mentioned above. 
One group of prior works focuses on the bandwidth challenge, either alleviating bandwidth demand by designing dedicated cores that exploit graph sparsity\cite{HyGCN} or satisfying bandwidth demand by incorporating Process-In-Memory (PIM) in the accelerator\cite{gnn:accel:pim,ReGraphX}. 
However, most of these solutions lack flexibility in supporting more powerful yet complex GNNs, as their designs typically make strong assumptions about the targeted GNN models, limiting their applicability to a broader range of GNNs. 
Another group of works~\cite{EnGN,DAC:AutenT020} offers high flexibility for various GNNs but fails to address the bandwidth challenge, resulting in long-distance data movement problems and limited performance~\cite{ReGraphX}.

\subsection{Our Goal.}
The limitations of prior work motivate us to tackle the challenges of bandwidth requirement and model variety concurrently. 
Our approach explores generic and cross-stack optimizations for GNN computation, making few assumptions about GNN model structures. 
We integrate these optimizations into a single framework called \proj{}, which we will detail in the following sections.

\section{\proj{} Design}
\label{sec:design}

\subsection{\proj{} Overview}
\label{sec:overview}

\Fig{fig:workflow} illustrates the \proj{} workflow, which encompasses three core methods that form the backbone of \proj{}. 
Specifically, a GNN model $M$ written in a high-level language such as DGL or PyG is first compiled by our GNN compiler to process a graph $G$. 
This spans across the algorithm, software, and hardware layers.

\begin{algorithm}[t]
	\caption{\tbf{\tit{Partition-Level Operator Fusion}}}
	\label{alg:plof}
	
	\DontPrintSemicolon
	\SetKwInput{Input}{Input}
	\SetKwInput{Output}{Output}
	\SetKwFor{ParallelFor}{for}{do in parallel}{end}
	
	\Input{Dst. intervals \algit{I} and shards \algit{S} by DSW-GP.}
	\Output{None, updating the intervals \algit{I} in situ.}
	
	\For{{\normalfont each}~\algit{interval}~{\normalfont in}~\algit{I}~}{
		\algit{interval \leftarrow} \tit{\tbf{ScatterPhase}}\algit{(interval)};	\\
		\For{{\normalfont each}~\algit{shard}~{\normalfont in}~\algit{S[interval]}~}{
			\algit{interval \leftarrow} \tit{\tbf{GatherPhase}}\algit{(interval,shard)}; 
		} 
		\algit{interval \leftarrow} \tit{\tbf{ApplyPhase}}\algit{(interval)};
	}
\end{algorithm}

\subsection{Partition-Level Operator Fusion (PLOF)}
\label{sec:plof}

PLOF is co-designed with compiling-level operator fusion and dual-sliding-window-based graph processing style to eliminate redundant DRAM accesses between GNN operators. 
In contrast to prior operator fusion for Deep Neural Networks (DNNs), which only considers the program of the DNN model~\cite{tvm}, PLOF also takes the graph data structure into account to introduce a new fusion paradigm. 
However, due to the irregularity of the graph, determining which operators should be fused and at which abstraction level remains challenging.

To achieve PLOF's goal, we propose fusing operators that process vertices or edges at the graph interval and shard level. 
The GNN model is first divided into multiple phases, and the input graph is partitioned into intervals and shards. 
Then, all phases will iterate either the intervals or shards according to the DSW-GP to complete the GNN computation. 
In this scenario, the total off-chip memory access is roughly $n_p\times M$ instead of $n_o\times M$, where $n_p$ is the phase number, $n_o$ is the operator number, and $M$ is the off-chip memory access of one operator. 
Consequently, each phase can be regarded as a fused operator, which alleviates the high bandwidth requirement.

We employ a template program to separate a GNN model and iterate the graph, defining three \textit{phases}: \texttt{ScatterPhase}, \texttt{GatherPhase}, and \texttt{ApplyPhase}. 
The pseudo code is presented in \Alg{alg:plof}. 
The template is inspired by traditional graph processing~\cite{GAS}, where a graph analytic algorithm can be represented using a similar three-stage GAS programming model. 
Instead of the GAS model operating on a single vertex or edge, we batch them into vertex intervals and edge shards produced by DSW-GP to further improve locality and expose parallelism. 
A GNN model can thus be represented by phases of the template program operating on different parts of the graph. 
\Fig{fig:gc}-d shows an example of the PLOF program written in our ISA (\Sec{sec:isa}) corresponding to the high-level language in \Fig{fig:gc}-a.

\paragraph{Compiler Support.}
Assigning each operator in a GNN model to the appropriate phase is a non-trivial task due to the large number of operators involved. 
A GNN model typically consists of multiple layers, with each layer containing tens of operators, including GTRs. 
Consequently, it can be challenging for programmers to take advantage of the proposed operator fusion technique. 
However, the semantics of each operator can be inferred from specific operators within the model, offering an opportunity to automate the phase construction process through software support. 
To achieve this, we incorporate the process into our compiler, which will be detailed later in \Sec{sec:implementation:compiler}.

\begin{figure}[b]
	\centering
	\includegraphics*[trim=0.cm 0.cm 0.cm 0.cm,width=1\linewidth]{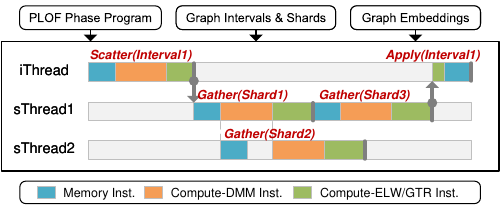}
	\vspace{-6pt}
	\caption{Executing PLOF phases using Shard-Level Multi-ThreadingMulti-Threading with 2 sThreads.}
	\label{fig:multithreading}
\end{figure}

\subsection{Shard-Level Multi-Threading (SLMT)}
\label{sec:slmt}

We propose SLMT to balance hardware flexibility and performance. 
In contrast, previous works~\cite{HyGCN} primarily focus on designing hardwired accelerators dedicated to specific GNN models. 
Although these designs achieve remarkable performance and efficiency, they suffer from limited flexibility. 
To address the variety challenge, we carefully trade off hardware flexibility and performance via SLMT.

The SLMT exploits shard-level parallelism, a new parallelism type dedicated to graph-related computation. 
We can parallelize shards since most operators in GatherPhase for shard processing have minimal dependency between shards, except for the GatherOp. 
Additionally, the shard is an ideal abstraction for parallelization due to its suitable size for the targeted memory and adjustability.

To implement SLMT, we construct a GNN accelerator with simultaneous multi-threading~\cite{smt}. The approach allows the hardware to automatically schedule different shards to issue their GatherPhase instructions, enabling those shards to utilize different hardware resources, such as functional units and bandwidth, simultaneously. 
\Fig{fig:multithreading} demonstrates this process, where multiple shards are processed concurrently by two shard-threads (sThreads). 
As shown in the figure, SLMT optimizes hardware resource utilization during shard processing, which is the central aspect of the overall GNN computation. 
We will describe the architectural details later in \Sec{sec:ga}.

\begin{figure}[b]
	\centering
	\includegraphics*[trim=0.1cm 0cm 0.1cm 0cm,width=1\linewidth]{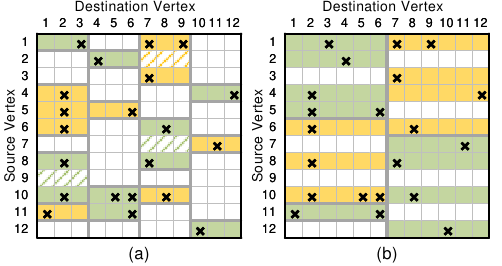}
	\vspace{-6pt}
	\caption{Graph shards produced by (a) prior graph partitioning with sparsity elimination~\cite{HyGCN} and (b) our fine-grained graph partitioning.
	}
	\label{fig:shard-construct}
\end{figure}

\subsection{Fine-Grained Graph Partitioning (FGGP)}
\label{sec:fggp}

FGGP is proposed to reduce redundant and unnecessary data transfers for partitioning-based GNN computation by exploiting graph sparsity. 
Generally, FGGP achieves this goal in two ways. 
First, FGGP can significantly increase the destination interval size (also the shard width) under memory constraints. 
We use \Fig{fig:shard-construct}-a as an example to illustrate the benefit. 
Initially, the source vertex 5 would be loaded twice under the first and second destination intervals. 
By increasing the interval size from 3 to 6, the source vertex 5 will only be loaded once, saving bandwidth. 
In this context, we identify the multiple loads of a single source vertex as redundant data transfer. 
The second aspect of FGGP is to skip unused source vertices for each shard, eliminating useless data transfers.

The core of FGGP lies in generating shards at the level of individual edges and source vertices during graph partitioning. 
To generate a shard, rather than forming a list of consecutive source vertices and simply assuming each source is fully connected to the vertices in the destination interval, we propose dynamically adding each edge with its associated source vertex to the shard until it is full. 
As a result, the source vertex lists of our shards can be discontinuous, as shown in \Fig{fig:shard-construct}-b. 
This method not only skips unused sources but also decouples the interval size from the memory constraint. 
Therefore, we can specify an interval size that far exceeds the memory size available for storing the shard.

We implement a graph partitioner to realize FGGP. 
To generate shards, the partitioner requires both the adjacency matrix of the input graph and the data dimensions from the compiler. 
We will discuss the graph partitioner in more detail later in \Sec{sec:gp}.

\section{\proj{} Implementation}
\label{sec:implementation}

To actualize the proposed methods, we develop a GNN accelerator framework comprising the GNN Accelerator (GA), GNN Compiler (GC), and Graph Partitioner (GP). 
Furthermore, we design an instruction set architecture (ISA) to serve as the interface connecting these three components of our framework.

In particular, the GA features a versatile instruction-driven pipeline equipped with various domain-specific functional units. 
Additionally, the GA employs SLMT to automatically pipeline GNN execution, ensuring efficient execution across a wide range of GNN models. 
The GC is devised to generate ISA code in the form of PLOF phases, accepting arbitrary GNN models written in high-level frameworks (e.g., DGL~\cite{dgl}, PyG~\cite{pyg}) as input, automatically mapping them into PLOF phases, and generating the corresponding ISA code. 
The GP employs FGGP to create denser shards from the input graph in accordance with the GA specification and GNN model information derived from the GC. 
We delve into the details of our developed framework in the following sections.

{\renewcommand{\arraystretch}{1.2}
	\begin{table}[t]
		\caption{Example of \proj{} instructions.}
		\label{tbl:ir}
		\centering\selectfont
		\resizebox{1\linewidth}{!}{
			\begin{tabular}{clc}
				\toprule
				\multicolumn{1}{c}{\textbf{Type}}    
				& \multicolumn{1}{c}{\textbf{Opname}}                                                                                                             
				& \multicolumn{1}{c}{\textbf{Operand}}                                                                      
				\\ \midrule
				\multicolumn{1}{c}{Compute}                
				& \begin{tabular}[c]{@{}l@{}}ELW (\ttt{ADD}, \ttt{MUL}, \ttt{RELU}), DMM (\ttt{GEMM}),\\ GTR (\ttt{GTHR.SUM.F}, \ttt{SCTR.F}, \ttt{SCTR.B}) \end{tabular}  
				& \makecell[r]{\multirow{3}{*}{\begin{tabular}[c]{@{}c@{}}Dimensions, \\Memory\\Symbols\end{tabular}}}
				\\ \cmidrule{1-2} 
				\multicolumn{1}{c}{Memory}                
				& \begin{tabular}[c]{@{}l@{}}Load/Store Src/Dst (\ttt{LD.D}, \ttt{LD.S}, \ttt{ST.D})\end{tabular}                                
				& 
				\\ \bottomrule
			\end{tabular}
		}
	\end{table}
}

\subsection{Instruction Set Architecture (ISA)}
\label{sec:isa}

\begin{figure}[t]
	\centering
	\includegraphics*[trim=0.cm 0.cm 0.cm 0.cm,width=1.\linewidth]{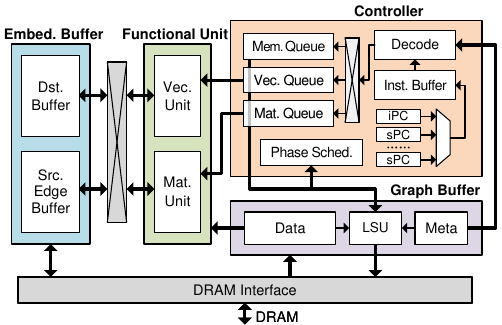}
	\caption{The architecture of \proj{} GNN Accelerator.}
	\label{fig:arch}
\end{figure}

We first introduce the ISA of \proj{}, which serves as the foundation for our GA, GC, and GP. 
\Tbl{tbl:ir} presents an example of the instructions, while \Fig{fig:arch} illustrates the GNN accelerator architecture.

Initially, we categorize the instructions in the ISA into two types: Compute and Memory. 
Each instruction type targets a specific architectural component within the hardware. 
Compute instructions are further classified into three sub-types, corresponding to the three primitive operator types in GNNs (\Sec{sec:background}). 
Generally, compute instructions are directed to the functional unit depicted in \Fig{fig:arch}. 
Memory instructions, on the other hand, handle data transfer of vertices and edges between the on-chip embedding buffer and off-chip DRAM and are issued to the load-store unit (LSU) shown in \Fig{fig:arch}.

For each instruction, we define three fields: opname, data-dimension, and memory-symbol. 
The opname and data-dimension fields specify the targeted operator and the associated parameters of the input and output data. 
To apply a single program to multiple intervals and shards with varying sizes, we also establish a set of macros representing the parameters of intervals and shards in the second field. 
For instance, we use \textit{E} for the number of edges in the current shard. 
These macros should be decoded at runtime by the hardware controller. 
The third field, memory-symbol, indicates the memory addresses of the input and output data. 
We define memory-symbols as another kind of macro, offering three symbol types—\textit{D} (destination), \textit{S} (source), and \textit{E} (edge)—to denote the data types of input and output, which may correspond to different hardware stores. 
To obtain the actual addresses, the hardware controller calculates the address at runtime, also accounting for the varying sizes of intervals and shards.

\begin{figure*}[t]
	\centering
	\includegraphics*[trim=0.cm 0.cm 0.cm 0.cm,width=1\linewidth]{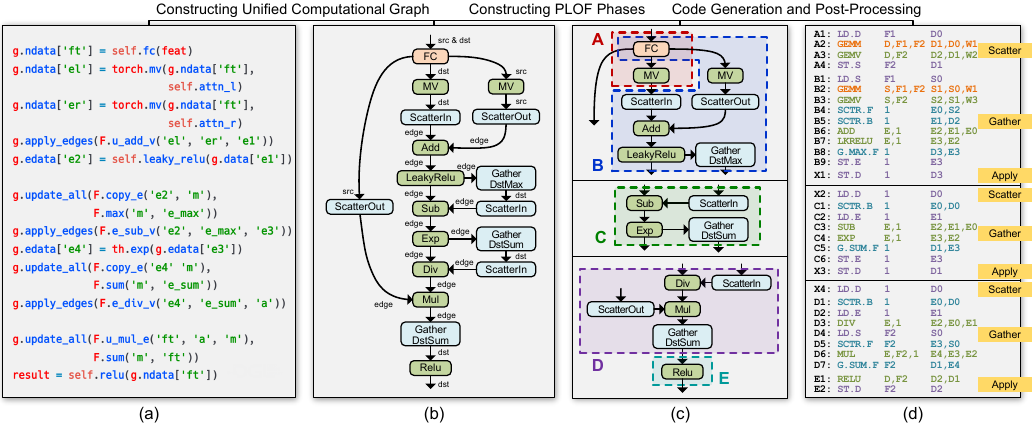}
	\caption{The workflow of \proj{} GNN Compiler with an example DGL program.}
	\label{fig:gc}
\end{figure*}

\subsection{GNN Accelerator (GA)}
\label{sec:ga}

The GA is an instruction-driven platform designed to perform various GNN computations as expressed by the ISA. 
To address the challenge of supporting diverse operator numbers and orders, we utilize the SLMT as the core of GA. 
Based on SLMT, we achieve a more flexible design where different functional units connect to various buffers in parallel, enabling versatile data access. 
Our GA design is illustrated in \Fig{fig:arch}.

\subsubsection{Functional Unit}
The GA comprises two types of dedicated functional units for GNN primitive operators (\Sec{subsec:background:gnn}). 
The first, called a \textit{vector unit (VU)}, includes multiple cores operating in the Single-Instruction-Multiple-Data (SIMD) paradigm. 
It targets the lightweight ELW and GTR operators. 
For ELW, each core operates independently on different embeddings. 
For GTR, each core is responsible for one destination vertex in GatherOp or one edge in ScatterOp. 
The second, a \textit{matrix unit (MU)}, is an output-stationary systolic multiply-accumulate (MAC) array targeting compute-intensive DMM operators for increased efficiency. 
In summary, this functional unit covers all three operator types in GNNs.

\subsubsection{Controller}
The Controller implements the SLMT as described in \Sec{sec:slmt}. 
It maintains multiple program counters (PCs), each corresponding to a thread. 
To simplify control flow, we employ one \textit{interval thread (iThread)} and multiple \textit{shard threads (sThreads)}. 
The iThread executes only the ScatterPhase and ApplyPhase, processing destination vertices in intervals, while sThreads execute only the GatherPhase, processing source vertices and edges in shards. 
For each instruction fetched by a PC, the controller decodes it and enqueues it into a queue specified by its instruction types. 
Each queue will dequeue and issue an instruction when the targeted component is ready.

To adhere to the execution logic in \Alg{alg:plof}, we introduce a Phase Scheduler (PS) to manage phase switching comprehensively. 
Whenever a PC reaches the end of a phase, the PS is activated to determine the next phase for the PC or to pause or resume a thread by accessing the shard metadata in the Graph Buffer. 
For instance, when an sThread completes a shard, the PS resets its PC and assigns it another shard if any are available within the interval. 
If no shards are left within this interval, the PS pauses the sThread. 
If all sThreads are paused, the PS resumes the iThread and sets its PC to ApplyPhase.

\subsubsection{Embedding Buffer}
We employ two on-chip scratchpad memory (SPM) pieces as Embedding Buffers to store vertex and edge embeddings and intermediate data. 
These buffers connect in parallel to the functional units through a crossbar, enhancing data access flexibility and allowing for arbitrary operator numbers and orders to be executed on the GA. 
One embedding buffer, called DstBuffer, stores data of destination vertices in intervals, corresponding to memory-symbol D in the ISA. 
The other embedding buffer, named SrcEdgeBuffer, stores data of edges and source vertices in shards, corresponding to memory-symbol S and E in the ISA. 
As the buffer holds various data for intervals and shards, we develop a compiler (\Sec{sec:implementation:compiler}) to effectively manage and access this buffer.

To supply data for the SLMT, we logically divide the SrcEdgeBuffer into $num\_sthread$ parts to hold multiple shards simultaneously. 
Each sThread privately accesses shard data in the SrcEdgeBuffer using different base addresses, while all threads share access to interval data in the DstBuffer.

\subsubsection{Graph Buffer}
The Graph Buffer is composed of three parts: MetaBuffer, DataBuffer, and LSU. 
The DataBuffer stores vectors representing connections between vertices and edges of a shard, using Coordinate Format (COO) to support GTR operators performed by the Functional Unit. 
The MetaBuffer stores scalars indicating current shard sizes, including the number of source vertices and edges. 
Both DataBuffer and MetaBuffer are designed to hold multiple shards simultaneously, supporting the SLMT. 
The LSU handles shard and embedding transfers, receiving memory instructions from the controller, translating them into low-level transactions using shard and interval data in the DataBuffer, and sending the transactions to the downstream DRAM interface.

We implement a straightforward prefetch mechanism to supply data to graph buffers. 
A 1-bit flag is assigned to each shard, indicating whether the data requires an update. 
If the bit is set to $1$, the current data is deemed outdated, prompting the LSU to automatically load the subsequent shard into the buffer and change the bit to $0$. 
The system remains idle until the phase scheduler in the controller switches the bit back to $1$ due to phase changes.

\subsection{PLOF Compiler}
\label{sec:implementation:compiler}

We develop the Graph Compiler (GC) to automatically map a GNN model, written in a high-level framework, to the PLOF template program. 
\Fig{fig:gc} illustrates the workflow of GC. 
The entire compilation process consists of three steps, as described below.

\subsubsection{Constructing Unified Computational Graph}
\label{sec:gc:1}
In this phase, GC extracts a unified computational graph from high-level frameworks. 
This unified graph replaces framework-specific graph operators—such as \texttt{apply\_edges} and \texttt{update\_all} in DGL, and \texttt{scatter} in PyG—with more generic GTR operators, as presented in \Sec{sec:background}.

\subsubsection{Constructing PLOF Phases}
\label{sec:gc:2}
GC decomposes the unified computational graph into \textit{groups} of PLOF phases, as outlined in \Sec{sec:plof}. 
Specifically, GC traverses the unified graph from each GTR operator along both in-edges and out-edges, respectively, until encountering another GTR. 
During traversals, GC labels each visited edge with \textit{src}, \textit{dst}, or \textit{edge}, depending on the starting GTR operator type. 
Subsequently, GC performs reverse topology sorting on the labeled unified graph, cutting the foremost edge of each successive edge block—marked with dst labels—and other corresponding unvisited edges.

Next, GC determines the phase to which each operator belongs. 
This process also relies on the data labels marked earlier. 
Specifically, GC executes reverse topology sorting, during which it appends visited operators to ApplyPhase until encountering an operator whose in-edge lacks a dst label. 
Then, GC appends visited operators to GatherPhase until encountering an operator whose out-edge does not have an edge label. 
Finally, GC appends all remaining operators to ScatterPhase.

\subsubsection{Code Generation and Post-Processing}

GC generates compute and memory instructions for operators in accordance with the ISA. 
Specifically, GC first creates one compute instruction for each operator. 
The opname and data-dimension are directly derived from the current operator, while the memory-symbols are jointly determined by the current and adjacent operators. 
For the destination memory-symbol, GC establishes its type based on the operator label marked in \Sec{sec:gc:2} and assigns a new number to the type. 
The source memory-symbols, therefore, are the destination symbols of its depending instructions. 
GC also inserts corresponding memory instructions into the phases if the input or output memory-symbols are not produced or consumed in this PLOF phases.

After instruction generation, GC performs memory-symbol liveness analysis to manage the on-chip buffer. 
GC first calculates the size of each symbol and then merges two symbols of the same size if the former is no longer in use. 
Consequently, more useful data can be stored on-chip, resulting in better buffer utilization. 
At this stage, GC completes code generation for GNNs from high-level frameworks.

During execution, GC produces parameters for graph partitioning. 
Specifically, GC calculates two parameters: $\textit{dim\_src}$ and $\textit{dim\_edge}$. 
The former represents the total data-dimensions of all source vertex memory-symbols in each GatherPhase, while the latter signifies the total data-dimensions of all edge memory-symbols. 
These parameters are determined by accumulating the data-dimensions of corresponding symbols. 
These parameters allow the downstream GP to perform FGGP, as detailed in the following \Sec{sec:gp}.

\begin{algorithm}[t]
	\caption{\tbf{\tit{DSW-GP with proposed FGGP}}}
	\label{alg:gp:fggp}
	
	\DontPrintSemicolon
	\SetKwInput{Input}{Input}
	\SetKwInput{Output}{Output}
	\Input{\algit{G}, input graph; \algit{M}, targeted GNN model.}
	\Output{\algit{S}, the resulted shards.}
	
	\algit{I \leftarrow} \tit{\tbf{initIntervals}}\algit{(G,M)};\\
	\For{{\normalfont each}~\algit{i}~{\normalfont in}~\algit{I}~}{
		\algit{s \leftarrow} \tit{\tbf{initShard}}\algit{(i)};\\
		\algit{srcPtr} $\leftarrow 0$;\\
		\While{\algit{srcPtr < G.vTotal}~} {
			\algit{dstList\leftarrow} \tit{\tbf{\blue{acquireNeiList}}}\algit{(G,i,srcPtr)};\\
			\If{\algit{dstList.size}~$>0$~} {
				\If{\tit{\tbf{\blue{probeShardSize}}}\algit{(s,dstList,M)}~} {
					\tit{\tbf{finalizeShard}}\algit{(i,s)};\\
					\algit{s \leftarrow} \tit{\tbf{initShard}}\algit{(i)};\\
				}
				\tit{\tbf{\blue{appendShardSource}}}\algit{(s,srcPtr,dstList)};\\
			}
			\algit{srcPtr \leftarrow srcPtr}$+1$;\\ 
		}
		\tit{\tbf{finalizeShard}}\algit{(i,s)};\\
	}
	\tit{\tbf{finalizeInterval}}\algit{(i)};\\
\end{algorithm}

\subsection{Graph Partitioner (GP)}
\label{sec:gp}

GP implement our FGGP based on DSW-GP. 
\Alg{alg:gp:fggp} presents the pseudocode for our GP. 
The primary distinction between our GP and the original DSW-GP, as shown in \Alg{alg:gp:bg}, lies in the inner loop, where we iterate through source vertices for each interval. 
For every source, GP first executes \textit{acquireNeiList} to obtain the \textit{dstList} of adjacent destination vertices, along with the associated edge under the current interval. 
If dstList is empty, GP directly skips the source; 
otherwise, GP performs \textit{probeShardSize} to check whether there is extra space for the source and its associated edges in the shard, based on the following rule:
\begin{equation}
	\begin{split}
		\textit{num\_src} \times \textit{dim\_src} + \textit{num\_edge} & \times \textit{dim\_edge} \\
		& \leq \frac{\textit{mem\_capacity}}{\textit{num\_sThread}},
	\end{split}
	\label{equ:gp}
\end{equation}
where $num\_src$ and $num\_edge$ represent the numbers of source vertices and edges of a shard, $dim\_src$ and $dim\_edge$ denote the total dimensions of all source vertex and edge memory-symbols in the generated phases from GC, and $num\_sthread$ is the number of sThreads running on GA. 
If \Equ{equ:gp} is satisfied, GP performs \textit{appendSource} to include the current source vertex and its associated edges into the shard. 
Otherwise, GP finalizes the current shard and initializes a new one before appending the source vertex.

{\renewcommand{\arraystretch}{1.2}
	\begin{table}[t]
		\caption{System configurations.}\label{tbl:baseline}
		\centering
		\resizebox{0.99\linewidth}{!}{
			\begin{tabular}{cccc}
				\toprule
				& \begin{tabular}[c]{@{}c@{}}\textbf{Computing}\\\textbf{Unit} \end{tabular}
				& \begin{tabular}[c]{@{}c@{}}\textbf{On-Chip}\\\textbf{Memory} \end{tabular}
				& \begin{tabular}[c]{@{}c@{}}\textbf{Off-Chip}\\\textbf{Memory} \end{tabular}
				\\ \midrule       
				
				\begin{tabular}[c]{@{}c@{}}\textbf{V100}\\\textbf{GPU}\end{tabular} 
				& \begin{tabular}[c]{@{}c@{}}80$\times$SIMT64 cores\\@1.25GHz\end{tabular} 
				& Totally 34MB 
				& \begin{tabular}[c]{@{}c@{}}900GB/s\\HBM-2\end{tabular}    
				\\ \midrule  
				\textbf{HyGCN}
				& \begin{tabular}[c]{@{}c@{}}16$\times$SIMD32 cores,\\8$\times$4$\times$128 systolic\\MAC array @1GHz\end{tabular}  
				& \begin{tabular}[c]{@{}c@{}}128KB(Input), 2MB(Edge),\\2MB(Weight), 4MB(Output),\\8MB(Aggregation)\end{tabular}
				& \begin{tabular}[c]{@{}c@{}}256GB/s\\HBM-1\end{tabular}
				\\ \midrule  
				\begin{tabular}[c]{@{}c@{}}\textsc{\textbf{Switch-}}\\ \textsc{\textbf{Blade}}\end{tabular}
				& \begin{tabular}[c]{@{}c@{}}16$\times$SIMD32 cores,\\32$\times$128 systolic\\MAC array @1GHz\end{tabular}  
				& \begin{tabular}[c]{@{}c@{}}8MB(DB), 1MB(SEB),\\2MB(Weight), 128KB(GB)\end{tabular}
				& \begin{tabular}[c]{@{}c@{}}256GB/s\\HBM-1\end{tabular}  
				\\ \bottomrule
			\end{tabular}      
		}
	\end{table}
}

\section{Methodology}
\label{sec:methodology}

\paragraph{Simulation Method.}
We implement, synthesize, and validate the \proj{} components using Verilog HDL with Synopsys Design Compiler and TSMC 28~nm standard cell library at 1~GHz.
Based on the synthesis, we construct a C++-based cycle-level simulator with aligned latency for end-to-end \proj{} evaluation. 
We integrate Ramulator~\cite{Ramulator} to obtain accurate latency for off-chip memory behaviors. 
Our simulator is validated against DGL built-in models~\cite{dgl} to ensure correct functionality. 
For on-chip scratchpad memory, we use Synopsys Memory Compiler to estimate area and power. 
We measure HBM access energy at 7 pJ/bit~\cite{GraphDynS}.

{\renewcommand{\arraystretch}{1.2}
	\begin{table}[t]
		\vspace*{-10pt}
		\caption{Graph datasets~\cite{gunrock-dataset} for evaluation.}
		\label{tbl:benchmark:dataset}
		\vspace{-6pt}
		\centering
		\resizebox{\linewidth}{!}{
			
			\begin{tabular}{cccc}\toprule
				\textbf{Dataset} & \textbf{Vertex\#} & \textbf{Edge\#} & \textbf{Description} \\
				\midrule
				ak2010 (AK)           & 45,293     & 108,549    & Redistrict Set            \\
				coAuthorsDBLP (AD)    & 299,068    & 977,676    & Citation Networks         \\  
				hollywood (HW)   & 1,139,905  & 57,515,616 & Collaboration Networks    \\ 
				cit-Patents (CP)      & 3,774,768  & 16,518,948 & Patent Networks           \\  
				soc-LiveJournal (SL) & 4,847,571  & 43,369,619 & Social Networks           \\
				\bottomrule             
			\end{tabular}
		}
	\end{table}
}

\paragraph{Benchmark Datasets and Models.}
We select five graphs from real-world workloads~\cite{gunrock-dataset} with varying graph sizes and sparsity levels, detailed in \Tbl{tbl:benchmark:dataset}. 
Additionally, we choose four popular and diverse GNN models~\cite{gnn-benchmark} with mathematical expressions shown in \Tbl{tbl:gnn-example}. 
For each model, we stack two identical layers with a dimension of 128 for input, hidden, and output embeddings for simplicity.

\paragraph{Baselines.}
We compare \proj{} with a general-purpose processor and a GNN accelerator. 
For the general-purpose processor, we use the DGL-0.7~\cite{dgl} library running on an NVIDIA V100 GPU\cite{v100} with 32~GB memory and an Intel Xeon E5-2630 v4 with 256~GB memory. 
For the GNN accelerator, we reproduce HyGCN, a state-of-the-art accelerator specifically designed for the GCN model, and compare its performance against \proj{} under the GCN. 
\Tbl{tbl:baseline} summarizes the configuration details. 
We set three sThreads for our shard-level multi-threading to match the three types of hardware resources: VU, MU, and bandwidth.

\section{Evaluation Results}
\label{sec:eval}

\begin{figure*}[t]
	\centering
	\includegraphics*[width=\linewidth]{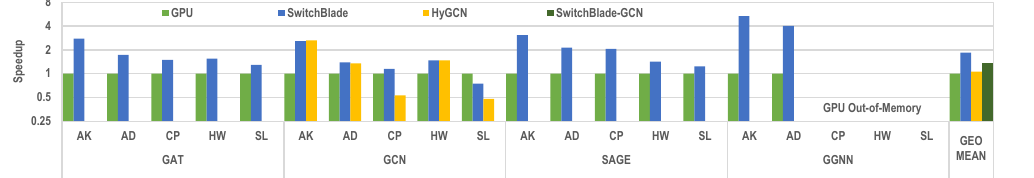}\\
	\vspace{-2pt}
	\caption{Speedup over the baseline V100 GPU.}
	\label{fig:eval:overall:latency}
\end{figure*}

\begin{figure*}[t]
	\centering
	\includegraphics*[width=\linewidth]{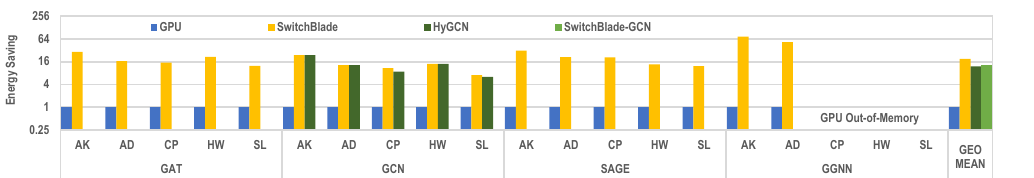}\\
	\vspace{-2pt}
	\caption{Energy saving over the baseline V100 GPU.}
	\label{fig:eval:overall:energy}
\end{figure*}

\subsection{Overall Results}
\label{subsec:eval:overall}

\paragraph{Latency.}
\Fig{fig:eval:overall:latency} compares the latency of different platforms. 
On average, \proj{} achieves a $1.85\times$ speedup over the baseline V100 GPU. 
\proj{} exhibits higher speedup on GAT, SAGE, and GGNN than GCN due to the presence of more operators in these models, providing greater opportunities for operator fusion and multi-thread scheduling. 
Moreover, \proj{} attains a $1.28\times$ speedup over HyGCN on GCN workloads, which is notable considering HyGCN deploys a hardwired pipeline tailored for GCN.

\paragraph{Energy.}
To ensure a fair comparison, we convert the results from 28nm to 12nm~\cite{convert}. 
\Fig{fig:eval:overall:energy} displays the results. 
\proj{} achieves an average $19.03\times$ energy saving over the baseline V100 GPU and $0.82\times$ over HyGCN. 
The high energy efficiency compared to the GPU is attributed to reduced on-chip data access and efficient domain-specific functional units in the \proj{} accelerator. 
In comparison to HyGCN, \proj{} also holds a slight advantage due to its simpler MU micro-architecture.

\paragraph{Area and Power.}
\Tbl{tbl:area-power} summarizes the area and power of \proj{} components under the TSMC 28~nm standard library. 
The total area and power of \proj{} are $28.25~mm^2$ and $6.06~W$, respectively, accounting for only $3.47\%$ and $2.43\%$ of the baseline V100 GPU with $815~mm^2$ and $250~W$ under the 12~nm technology node. 
Among the components, the SRAM-based SPM (SEB, DB, and TB) consumes the majority, representing $76\%$ and $58\%$ of the total area and power, respectively.


{\renewcommand{\arraystretch}{1.1}
	\begin{table}[t]
		\caption{Area and power breakdown of \proj{}.}
		\label{tbl:area-power}
		\centering
		\resizebox{0.9\linewidth}{!}{
			\begin{tabular}{c|cccc|c}
				\toprule
				\textbf{TSMC 28nm}  & \textbf{MU} & \textbf{VU} & \textbf{CTRL} &  \textbf{RAM} & \textbf{Total} \\ \midrule
				\textbf{Area / \%}  & 15.46       & 6.37        & 2.11          &  76.06        & 28.25 mm2       \\ \midrule
				\textbf{Power / \%} & 24.02       & 14.95       & 2.66          &  58.38        & 6.06 W          \\ \bottomrule
			\end{tabular}
		}
		 \vspace{-2pt}
	\end{table}
}

In the following subsections, we evaluate the effectiveness and sensitivity of each individual method proposed in this paper.

\subsection{Partition-Level Operator Fusion}
\begin{figure}[t]
	\centering
	\includegraphics*[trim=0.cm 0.cm 0.cm 0.2cm,width=1\linewidth]{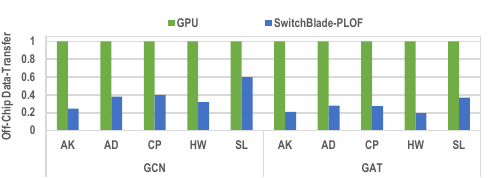}\\
	\caption{Normalized off-chip data-transfer with PLOF over GPU execution paradigm.}
	\label{fig:eval:plof}
\end{figure}

We assess the effectiveness of PLOF by measuring the reduction in total data transfer between on-chip and off-chip memory. 
\Fig{fig:eval:plof} presents the results. 
PLOF is effective for all types of GNN workloads and significantly reduces data transfer compared to the operator-by-operator execution paradigm on the GPU platform. 
This substantial reduction contributes to the speedup and energy reduction over the baseline V100 GPU.

\subsection{Shard-Level Multi-Threading}

\begin{figure}[t]
	\centering
	\begin{minipage}{0.48\linewidth}
		\centering
		\includegraphics*[width=\linewidth]{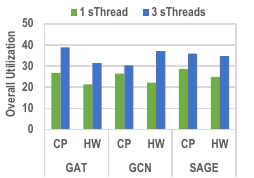}\\
		\caption{Overall utilization of hardware resources, including bandwidth and functional units with SLMT.}
		\label{fig:eval:slmt-utilization}
	\end{minipage}\hfill%
	\begin{minipage}{0.48\linewidth}
		\centering
		\includegraphics*[width=\linewidth]{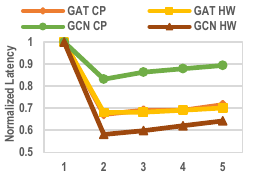}\\
		\caption{Execution latency using different SLMT sThread numbers normalized to 1 sThread.}
		\label{fig:eval:slmt-performance}
	\end{minipage}\hfill%
\end{figure}

\paragraph{Hardware Utilization.}
To evaluate the effectiveness of SLMT, we measure the overall hardware utilization by averaging the individual utilization of DRAM bandwidth, VU, and MU. 
\Fig{fig:eval:slmt-utilization} presents the results. 
\proj{} achieves higher overall utilization across all workloads when employing 3 SLMT sThreads rather than 1, which is regarded as SLMT turned off.

\paragraph{sThread Number.}
We also assess the performance and resource utilization under different concurrent sThread numbers. 
\Fig{fig:eval:slmt-performance} displays the results. 
Generally, the execution latency decreases initially and then increases, reaching optimal performance with two sThreads on each workload. 
The decrease in latency is expected as multiple sThreads can concurrently exercise different hardware execution units (VU, MU, and bandwidth), resulting in higher overall hardware utilization and performance. 
The reason for the increase in latency after two sThreads is the reduced data parallelism and hardware functional unit efficiency due to limited on-chip memory space for each sThread. 
Consequently, we observe minimal performance improvement with more than three sThreads based on the three types of hardware execution units.

\subsection{Fine-Grained Graph Partitioning}

\begin{figure}[t]
	\centering
	\begin{minipage}{0.48\linewidth}
		\centering
		\includegraphics*[width=\linewidth]{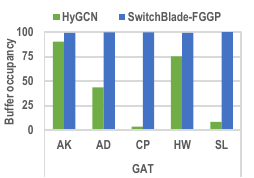}\\
		\caption{Occupancy rate of on-chip SPM with FGGP compared to HyGCN sparsity elimination.} 
		\label{fig:eval:fggp-buffer}
	\end{minipage}\hfill%
	\begin{minipage}{0.48\linewidth}
		\centering
		\includegraphics*[width=\linewidth]{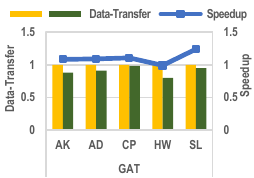}\\
		\caption{Total data transfer and speedup with larger SPM as same as HyGCN with FGGP.}
		\label{fig:eval:fggp-reuse}
	\end{minipage}\hfill%
\end{figure}

\paragraph{Data Density.}
To evaluate the effectiveness of mitigating the larger on-chip SPM requirement from SLMT, we first measure the average buffer occupancy rate for SEB and DB on different datasets, calculated as:
$$\textit{occupancy\_rate} = \frac{1}{W} \sum^i_{i\in[1,W]}\frac{\textit{data\_accessed}_i}{\textit{total\_buffer\_space}} $$
where $W$ is the total number of writes to the targeted buffer. 
\Fig{fig:eval:fggp-buffer} presents the results. 
The occupancy rate of \proj{} reaches nearly $99\%$ using FGGP, whereas prior graph partitioning with sparsity elimination in HyGCN only achieves a $44\%$ occupancy rate. 
As a result, \proj{} can attain similar performance using $31\%$ less on-chip SPM.

\paragraph{Data Reuse.}
We also measure the total data transfer and corresponding performance improvement from FGGP under the same on-chip SPM budget as HyGCN. 
\Fig{fig:eval:fggp-reuse} displays the results. 
By increasing the DB size from 8MB to 13MB, we achieve an additional $10\%$ data-transfer reduction through FGGP, leading to a further $1.1\times$ speedup. 
FGGP gains less speedup on the HW graph because HW is relatively dense, making the bandwidth less critical than functional units and rendering the FGGP optimization less effective.
\section{Related Work}\label{sec:related}

In the past three years, GNN hardware acceleration has become a prominent research topic. 
Some studies have achieved high performance and efficiency through model-specific optimizations. 
HyGCN~\cite{HyGCN} is the first to decouple the two operators in GCNs into stages, designing two dedicated compute engines to enable a two-stage pipeline execution. 
GReTA~\cite{GReTA} extends this concept with a multi-stage pipeline and bidirectional dataflow. 
GraphACT~\cite{GraphACT} and ReGraphX~\cite{ReGraphX} also incorporate CPU and 3D ReRAM techniques in their multi-stage acceleration. 
AWB-GCN~\cite{AWB-GCN} proposes a unified \ttt{SpMM} architecture for GCN's two operators, improving hardware utilization. 
GCNAX~\cite{GCNAX} presents a reconfigurable loop ordering and fusion design for GCN, while I-GCN~\cite{I-GCN} explores data locality with specific reorder mechanisms to enhance accelerator performance. 
These approaches achieve high performance and efficiency by making strong assumptions about the targeted GNN models and designing specialized hardware architectures. 
Although they can run other models that do not match their hardware designs by bypassing some stages, this increases data movement and subsequently reduces performance and efficiency.

Other studies focus on more flexible hardware or scheduling designs. 
EnGN~\cite{EnGN} introduces a unified SIMD architecture that adopts a ring-based dataflow for \ttt{SpMM} to improve hardware utilization. 
Auten et al.~\cite{DAC:AutenT020} connect all components with a crossbar switch in a hardware block. 
However, these works do not capture the generic GNN characteristics and suffer from long-distance data movement problems, leading to low performance and energy efficiency. 
In contrast, we make no assumptions about the targeted GNN models and propose three methods to enhance GNN accelerator performance. 
Our approach involves co-designing the software and hardware of the compiler, architecture, and graph partitioner as a whole.

\section{Conclusion}
\label{sec:conclusion}

In this paper, we present \proj{}, a comprehensive framework designed to address both variety and bandwidth challenges in GNN acceleration. 
Our approach achieves an average speedup of $1.85\times$ and energy savings of $19.03\times$ compared to GPUs, while also delivering performance on par with state-of-the-art model-specific GNN accelerators.
To accomplish these results, we introduce a partition-level operator fusion technique that significantly reduces the high bandwidth requirements associated with GNN execution. 
Furthermore, we propose a shard-level multi-threading approach to enhance the overall utilization of bandwidth and other functional units. 
Lastly, to tackle the increased on-chip memory contention caused by multi-threading, we present a fine-grained graph partitioning method for refining shard data. 
Importantly, all three methods are model-agnostic, making them suitable for a wide range of GNN models. 
Our experimental results demonstrate the effectiveness of the proposed methods in \proj{} for enhancing the performance of various GNN models and datasets.



\bibliographystyle{IEEEtranS}
\bibliography{refs}

\vspace{11pt}

\section{Biography Section}
\vspace{-20mm}
\begin{IEEEbiography}[\vspace{-15pt} {\includegraphics[width=1in,height=1.25in,clip,keepaspectratio]{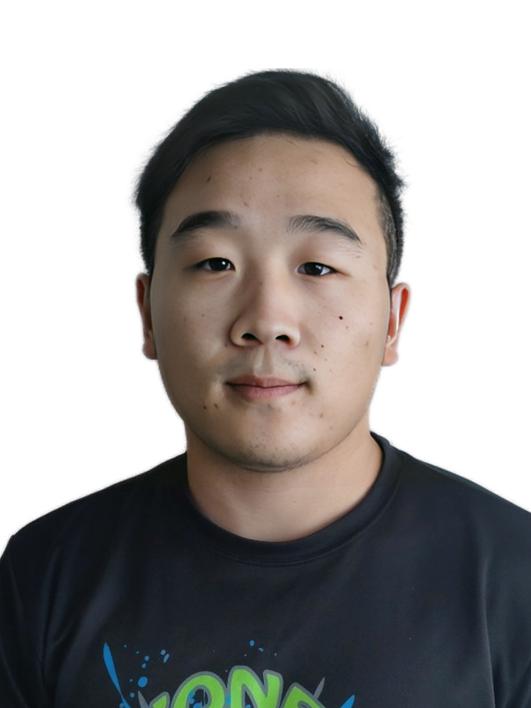}}]
        {Shuwen Lu} received the B.Sc. degree from Shanghai Jiao Tong University, China. He is currently a master candidate in computer science under the supervision of Dr. Jingwen Leng at the Department of Computer Science and Engineering of Shanghai Jiao Tong University, China. His research interests include GNN accelerating and high-performance computing.
\end{IEEEbiography}
\vspace{-15mm}
\begin{IEEEbiography}[{\vspace{-15pt}\includegraphics[width=1in,height=1.25in,clip,keepaspectratio]{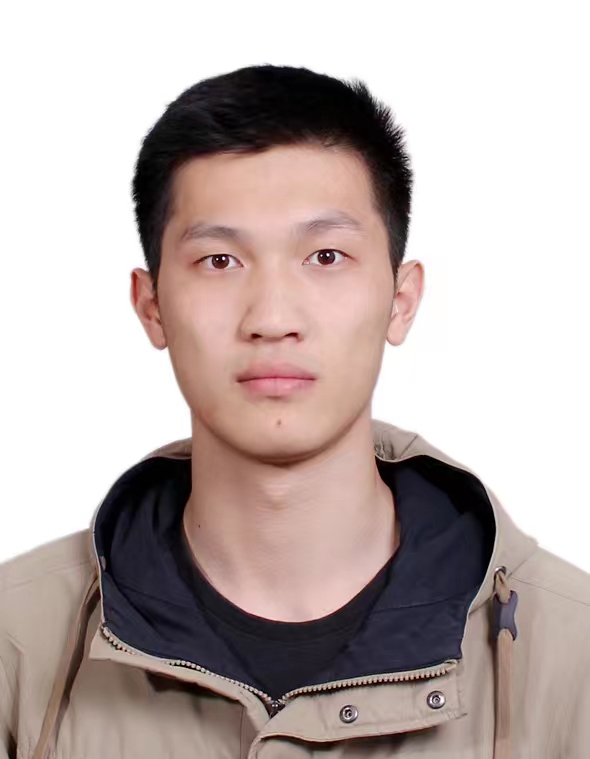}}]
        {Zhihui Zhang} received the B.E. degree from Beijing Jiaotong University, China, in 2019. He then pursued Ph.D. studies in computer science at the Department of Computer Science and Engineering of Shanghai Jiao Tong University, China. His research interests include GPU architecture, domain-specific architecture for graph neural networks, and architectural modeling and simulation. 
\end{IEEEbiography}
\vspace{-15mm}
\begin{IEEEbiography}[\vspace{-25pt} {\includegraphics[width=1in,height=1.25in,clip,keepaspectratio]{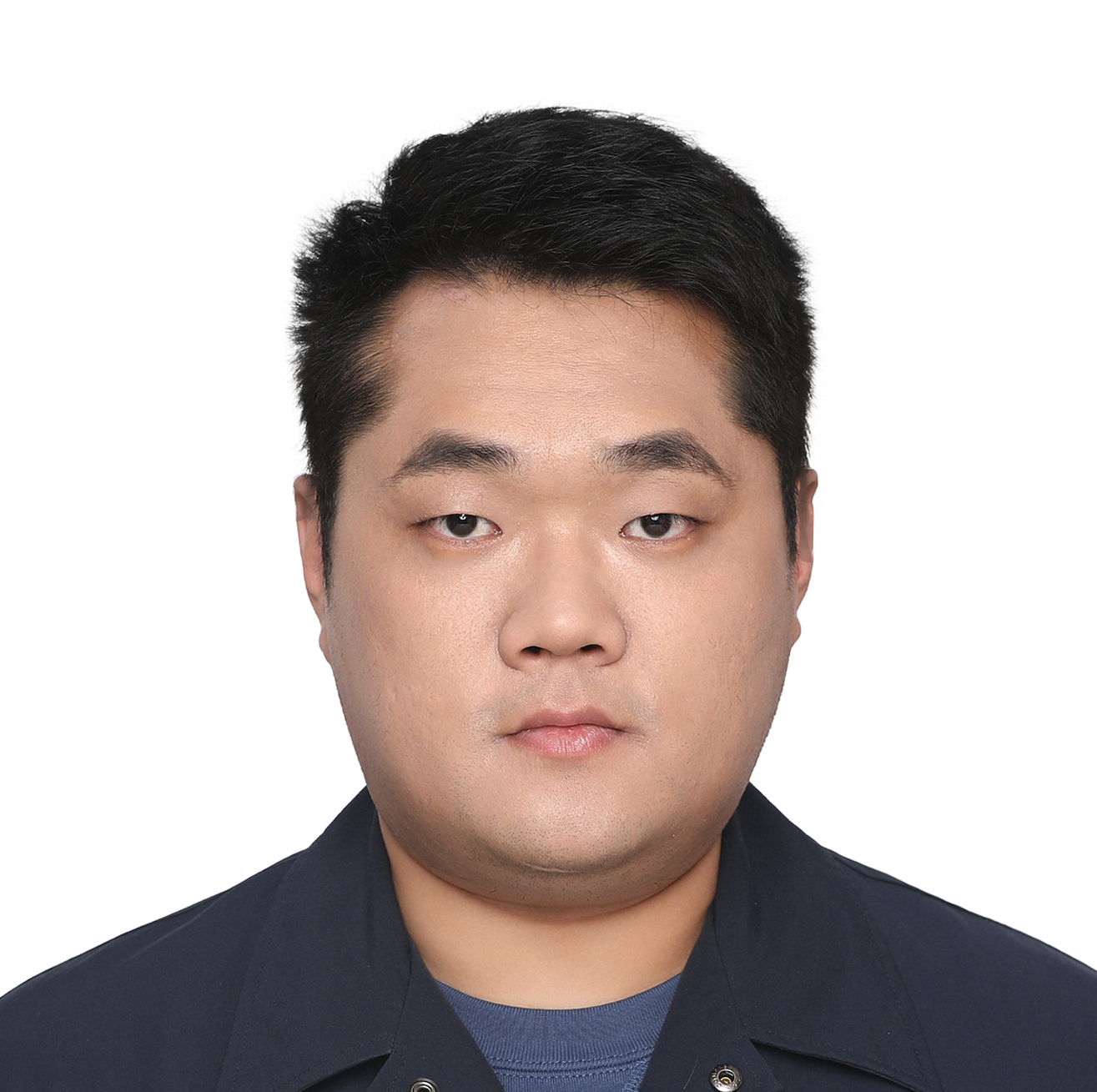}}]
        {Cong Guo} received the B.Sc. degree from Shenzhen University, China. He is currently a Ph.D. candidate in computer science under the supervision of Dr. Jingwen Leng at the Department of Computer Science and Engineering of Shanghai Jiao Tong University, China. His research interests include computer architecture, high-performance computing, and AI accelerator design.
\end{IEEEbiography}
\vspace{-15mm}
\begin{IEEEbiography}[\vspace{-5pt} {\includegraphics[width=1in,height=1.25in,clip,keepaspectratio]{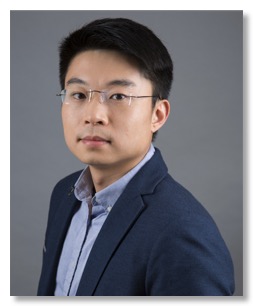}}]
        {Jingwen Leng} is a tenured Associate Professor in John Hopcroft Computer Science Center and Computer Science and Engineering Department at Shanghai Jiao Tong University. He received the Ph.D. degree from the University of Texas at Austin. He was the lead co-author for GPUWattch, one of the most widely used open-sourced GPU power model. He is currently interested at building intelligent and robust system for artificial intelligence.
\end{IEEEbiography}
\vspace{-15mm}
\begin{IEEEbiography}[\vspace{-5pt}{\includegraphics[width=1in,height=1.25in,clip,keepaspectratio]{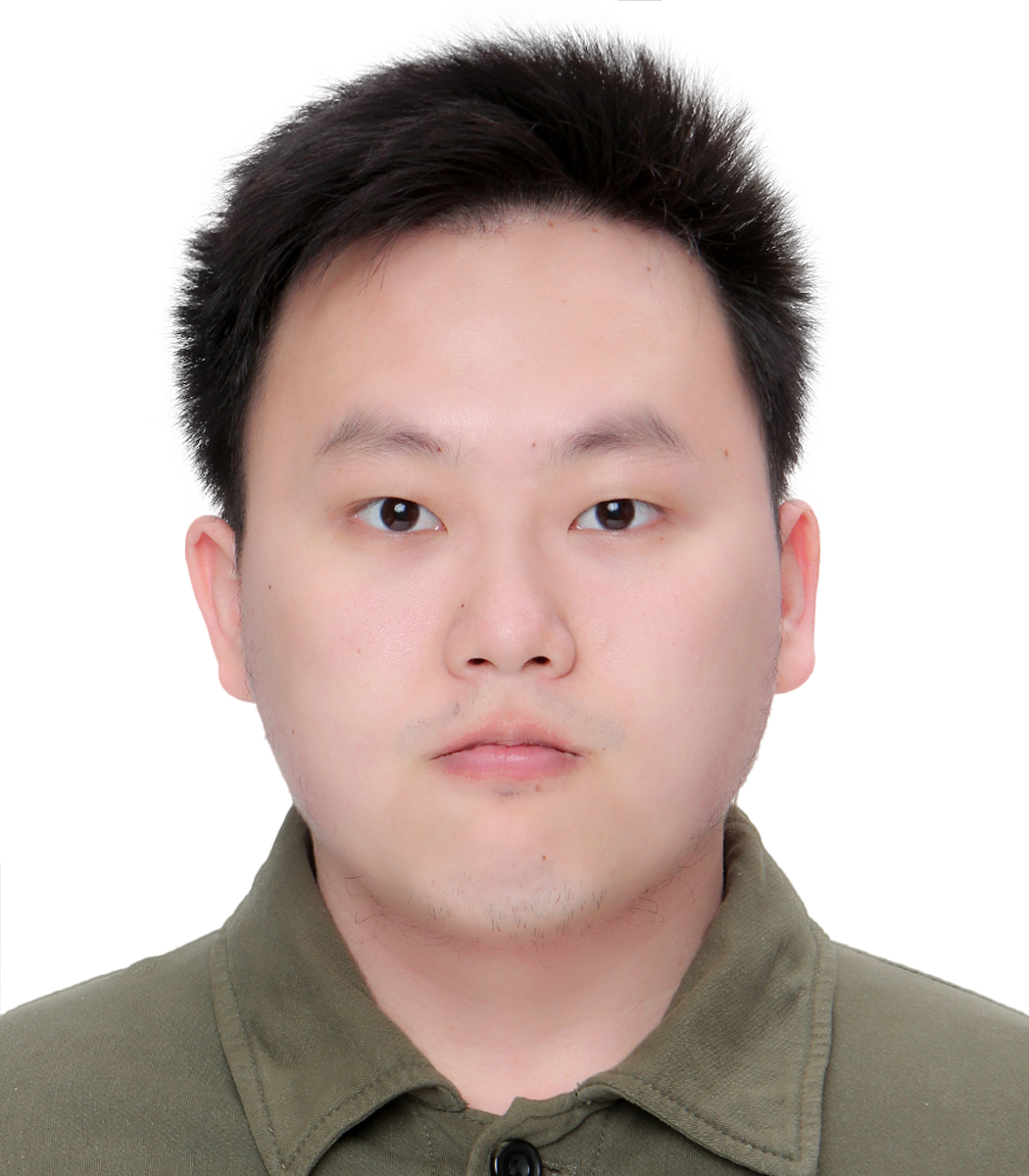}}]
        {Yangjie Zhou} received the B.S. degree in computer science from Huazhong University of Science and Technology, China. He is currently pursuing the Ph.D. degree under the supervision of Dr. Jingwen Leng at the Department of Computer Science and Engineering of Shanghai Jiao Tong University, China. Hs research interests include ML systems, high-performance computing, and computer architecture.
\end{IEEEbiography}
\vspace{-15mm}
\begin{IEEEbiography}[\vspace{-10pt}{\includegraphics[width=1in,height=1.25in,clip,keepaspectratio]{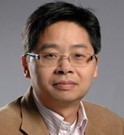}}]
        {Minyi Guo} (Fellow, IEEE) received the Ph.D. degree in computer science from the University of Tsukuba, Japan. He is currently Zhiyuan Chair professor in the Department of Computer Science and Engineering, Shanghai Jiao Tong University, China. His present research interests include parallel/distributed computing, compiler optimizations, embedded systems, pervasive computing, big data and cloud computing. He is now on the editorial board of IEEE Transactions on Parallel and Distributed Systems, IEEE Transactions on Cloud Computing and Journal of Parallel and Distributed Computing. Dr. Guo is a fellow of IEEE, and a fellow of CCF.
\end{IEEEbiography}

\end{document}